\documentstyle[12pt]{article}

\newcommand{\la}{\langle}
\newcommand{\ra}{\rangle}
\newcommand{\beq}{\begin{eqnarray}}
\newcommand{\eeq}{\end{eqnarray}}

\newcommand{\cl}{\centerline}

\begin{document}

\title{Isospin Symmetry Breaking in Hadrons and Nuclei
\footnote{Invited talk presented at
 the International Symposium on Spin-Isospin Responses
 and Weak Processes in Hadrons and Nuclei (March 8-10, 1994, Osaka, Japan).}
 }

\author{Tetsuo Hatsuda \\
\address{Institute of Physics, University of Tsukuba,
 Tsukuba, Ibaraki 305, Japan}}


\maketitle

\begin{abstract}

 A brief review of the isospin symmetry breaking in hadrons and nuclei
 is given with emphasis on  the
 $u-d$ quark mass difference.
  The  off-shell $\rho^0-\omega$ mixing is studied as a typical
 example of the symmetry breaking, and its relevance to
 the nuclear force and  nuclei is  discussed.

\end{abstract}

\section{Introduction}

In hadron and nuclear physics, the isospin symmetry has been  known to hold
 fairly well.  In the quark level, the symmetry is defined as
 a global vector rotation in $u-d$ isospace.
 It is also useful to define the charge symmetry as a 180$^\circ $ rotation
 around $y$ axis in the isospin space.  By this operation $R$, the
 $u$ and $d$ mass eigenstates transform as
\beq
R \mid u \ra = - \mid d \ra, \ \ \ \ R \mid d \ra =  \mid u \ra \ \ .
\eeq
Whenever charge symmetry breaking (CSB) occurs, so does the isospin symmetry
breaking.  Hereafter, we will focus on CSB for simplicity.

CSB has  two detectable origins:

\noindent
(i) Electromagnetic (EM) effect:  Since the electric charges of the $u$ quark
 and the $d$ quark are  different ($e_u =2/3\cdot e$, $e_d=-1/3\cdot e$),
 QED induces CSB of $O(\alpha)$.

\noindent
(ii) Quark mass difference:
 Since the current quark masses of the $u$ quark and the $d$ quark are
 different ($m_u(1{\rm GeV})\simeq 4 {\rm MeV}$,
$m_d(1{\rm GeV})\simeq 9 {\rm MeV}$ \cite{GLR}),
 the quark-mass term $H_m^{QCD}$ in the QCD hamiltonian
 induces CSB.  This is ease to see from
\beq
 H_{m}^{QCD} = {1 \over 2}(m_u+m_d) (\bar{u}u+\bar{d}d)
                + {1 \over 2}(m_u-m_d) (\bar{u}u-\bar{d}d)\ \ .
\eeq
The second term proportional to $m_u-m_d$ gives $O((m_d-m_u)/\Lambda_{QCD})$
 breaking of the charge symmetry.

 Since CSB is generally small ($1\%$ level or less),
 it is a good approximation to expand arbitrary physical quantity
 $A$ up to first order in $\alpha$ and $m_d-m_u$:
\beq
A(\alpha,m_d-m_u) \simeq
 A(0,0) + b \cdot \alpha + c \cdot {{m_d-m_u} \over \Lambda_{QCD}}\ \ .
\eeq

\subsection{CSB in hadrons}

Among various examples of CSB phenomena in hadron physics, the
following three are well known and relatively important:
 $\pi^{\pm}-\pi^0$ mass difference, proton-neutron mass difference and
 the $\rho^0-\omega$ mixing (See Table 1).
 The first one is a typical example of the EM dominated CSB: Since
 $\pi^{\pm}$ and $\pi^0$ have equal number
 of $u$ and $d$ quarks  on the average,
  $O(m_d-m_u)$ effect vanishes in the mass difference.
 This is the reason why the naive expectation ``charged particles should have
  larger inertia mass than neutral particles due to the photon cloud''
 holds in this case.  Theoretical estimate of the EM effect for the
 pion self-energy is
 consistent with this picture.
 On the other hand, the neutron is heavier than the proton in the real world,
 which indicates that there must be a sizable CSB from the
quark mass difference
 with opposite sign to EM effect.
 The recent measurement of the $e^+e^- \rightarrow \pi^+\pi^-$
 shows an unambiguous determination of the $\rho^0-\omega$ mixing with
negative sign \cite{exp,CB}.  The EM contribution due to
   $\rho^0 \rightarrow \gamma \rightarrow \omega$ is positive and small,
 thus  a relatively large and
 negative contribution from the quark mass difference is again necessary.
 The numbers in the 4th column of Table 1 are the expected $O(m_d-m_u)$
 contributions
 obtained by subtracting the theoretical EM effects (3rd column) from
the experimental numbers (2nd column).
The QCD sum rules as well as other effective models have been used to
  estimate the $O(m_d-m_u)$ effect and
 the results are consistent with those in Table 1.
(See e.g. \cite{SVZ79c,HHP}.)
 Thus we are rather  in a good shape to understand
 the origin of CSB in hadrons, although the first principle calculations on
 the lattice are still missing.

\begin{table}[hbt]
\newlength{\digitwidth} \settowidth{\digitwidth}{\rm 0}
\catcode`?=\active \def?{\kern\digitwidth}

\label{tab1}
\begin{tabular*}{\textwidth}{@{}l@{\extracolsep{\fill}}rrrr}
\hline
$A$  &  experiments  &  $O(\alpha)$   &  $O(m_d-m_u)$  \\
\hline
$m_{\pi^{\pm}}-m_{\pi^0}$ (MeV)  & $4.6$    & $4.6 \pm 0.1$ (theory)
 &   $0$        \\
$m_p-m_n$   (MeV)                & $-1.29$  & $0.76\pm 0.3$ (theory)
  & $\sim \ -2.05$  \\
$\rho^0$-$\omega$ mixing (MeV$^2$) & $-4520\pm600$  & $\sim 610$ (theory)
 & $\sim \ -5130 $ \\
\hline
\end{tabular*}

\vspace{0.3cm}
\noindent
Table 1.  CSB examples in hadron masses
 and the mixing. $\rho^0-\omega$ mixing
 is defined by the covariant matrix element
 $\la \rho^0 \mid H_{CSB} \mid \omega \ra $
 at the $\rho^0-\omega$ mass shell with
 $H_{CSB}$ being the second term in eq.(2).
 See \cite{GLR} for the details of the $O(\alpha )$ estimate.
 One should note here that $m_{\rho^0} \neq m_{\omega}$ even without CSB.

\end{table}

\subsection{CSB in nuclei}

What about CSB in nucleon-nucleon interactions and in nuclei?
There are, in fact, accumulated evidences that EM effect is not enough
 to explain the observed CSB.
 (In the following, we will put `` bar'' to express CSB with the
 EM effect subtracted.)
  For example, (i) the CSB in N-N scattering length such as
 $\bar{a}_{nn}-\bar{a}_{pp} = -1.5 \pm 0.5\ {\rm fm}$ \cite{CB,MIL}, (ii)
 binding energy differences in mirror nuclei such as
 $\bar{B}(^3 H)- \bar{B} (^3 He) = 80 \pm 24 {\rm keV}$ \cite{CB}
  and the Okamoto-Nolen-Schiffer  anomaly \cite{ONS,BI},
 and (iii) the asymmetry in the polarized $n-p$ scattering at zero-crossing
 angle $\Delta A (\theta_0)$ \cite{WTM}.
 See ref.\cite{MIL} for a
 more complete list of the nuclear CSB.

 As far as the two-body N-N interaction dominates in nuclei,
 all the above  quantities are related to the EM subtracted
 CSB in the elementary N-N interaction
 $\bar{V}_{NN}$;
\beq
\bar{V}_{nn} \neq \bar{V}_{pp} \neq \bar{V}_{np}\ \ .
\eeq
 Among various contributions to $\bar{V}_{NN}$,
 the exchange of the $\rho^0-\omega$ complex
  (the one in Fig. 1(b)) is recently claimed to play a crucial role
  in the above examples (i)-(iii).
 In the following sections, I will discuss whether this idea is
 plausible or not  on the basis of the recent work by
 Henley, Meissner, Krein and myself \cite{HHMK}.

\section{The off-shell $\rho^0-\omega$ mixing}

Let's start with the following hadronic correlation function
\beq
\Pi _{\mu \nu } (q^2)  \equiv  i \int d^4x \la {\rm T} \rho^0_{\mu}(x)
 \omega_{\nu}(0) \ra_0
 =  - (g_{\mu \nu} - {q_{\mu}q_{\nu} \over q^2})
{1 \over q^2-m_{\rho}^2}\cdot \Theta(q^2) \cdot
{1 \over q^2-m_{\omega}^2}\ \ .
\eeq
  The mixing matrix $\Theta$ is in general $q^2$ dependent.  It can be
 determined at least at $q^2 \simeq m_{\rho,\omega}^2$
 through the interference of
 $e^+e^- \rightarrow \rho^0 \rightarrow \pi^+ \pi^-$ with
 $e^+e^- \rightarrow \omega \rightarrow \rho^0 \rightarrow \pi^+ \pi^-$
 (Fig. 1(a))
  in the  $e^+e^- \rightarrow \pi^+ \pi^-$ experiment \cite{exp,CB}.
   The process in Fig. 1(a) is a $s$-channel exchange of the
 $\rho^0-\omega$ complex for $q^2 > 0$.
  On the other hand, in the elastic N-N scattering,
 the $\rho^0-\omega$ complex is  exchanged in the $t$-channel
  (Fig. 1(b)), thus the relevant $q^2$ is
 inevitably off-shell and space like ($q^2 < 0$).
  All the previous estimates of CSB in N-N force and in nuclei
 have however been done with an assumption that
  $\Theta(q^2)$ is $q^2$ independent
 in the wide range of the momentum
 $-m_{\rho,\omega}^2 < q^2  < m_{\rho,\omega}^2$.  Is this a valid
 assumption or not?
   This is the question first raised by Goldman, Henderson and
 Thomas \cite{GHT}.
  Subsequently, Meissner, Henley, Krein and myself have proved that
 $\Theta$ must be $q^2$ dependent in a model independent way,
 and  have  extracted the  dependence
 using the QCD sum rules \cite{HHMK}.

\vspace{7cm}

\noindent
Figure 1. (a) $s$-channel exchange of the $\rho^0-\omega$ complex
 in  $e^+e^- \rightarrow \pi^+ \pi^-$ experiment.
 (b) $t$-channel exchange in the N-N interaction.

\vspace{1.0cm}

\subsection{Why $\Theta$ must be $q^2$ dependent?}

The heuristic argument goes as follows.
 Let us concentrate on the non-EM part of $\Theta$ and
  start with an unsubtracted dispersion relation for
$\Pi(q^2) \equiv - \Pi_{\mu \mu}(q^2)/3$:
\beq
{\rm Re} \Pi(q^2) = {{\rm P} \over \pi} \int
 {{\rm Im}\Pi(s) \over {s-q^2}} ds \ \ .
\eeq
 Instead of $\Pi(q^2)$ here, one can make the same argument
 starting with the QCD correlation
$\Pi_{\mu \nu}^{QCD} \sim
 \la {\rm T} J_{\mu}^{\rho}(x) J_{\nu}^{\omega}(0) \ra_0$
 with $J_{\mu}^{\rho,\omega} =
(\bar{u} \gamma_{\mu}u \mp \bar{d} \gamma d)/2$.
 In that case, one can prove that the subtraction is not necessary for the
 non-EM part. Also one can take into account the higher resonances ($\rho'$
 and $\omega'$).  See \cite{HHMK} for details.

 Saturating the imaginary part in (6) by the $\rho^0$ and $\omega$ resonances
 as ${1 \over \pi}{\rm Im}\Pi(s) \equiv F_{\rho} \delta(s-m_{\rho}^2)
 - F_{\omega} \delta(s-m_{\omega}^2)$, one immediately arrives at
\beq
{\rm Re} \Pi(q^2) = {-F_{\rho} \over q^2-m_{\rho}^2}
                    + {F_{\omega} \over q^2 - m_{\omega}^2}
 = {1 \over q^2-m_{\rho}^2} \cdot \Theta(q^2) \cdot
                     {1 \over q^2 - m_{\omega}^2} ,
\eeq
with $\Theta$  being the non-EM part of the mixing matrix
\beq
{\Theta(q^2) \over \Theta(m^2)} = 1 + \lambda ({q^2 \over m^2} -1),
\eeq
where $m^2 = (m_{\rho}^2+m_{\omega}^2)/2$ and
  $\lambda = (F_{\rho}-F_{\omega})/(F_{\rho}+F_{\omega})\cdot
 (m_{\rho}^2+m_{\omega}^2)/(m_{\rho}^2-m_{\omega}^2)$.
One should note the following points in the above formula:

\noindent
(i) $F_{\rho} $ and $F_{\omega}$ are $O(m_d-m_u)$ quantities. However,
there is no reason to believe that $F_{\rho} = F_{\omega}$, since
they are the residues at different pole positions.
 Also note that $m_{\rho} \neq m_{\omega}$ even if there is no CSB.

\noindent
(ii) $\lambda$, which dictates the $q^2$ dependence
 of the $\rho^0-\omega$ mixing, is an $O(1)$ quantity. This suggests
 a large variation of $\Theta$ from the on-shell point to the space
 like points. However,
 in the previous applications of the $\rho^0-\omega$ mixing
to the nuclear  CSB,
 $\lambda$ is assumed to be zero with no  specific reason.

\noindent
(iii) It is unlikely that the relatively small EM effect (see the
 4th row of Table 1) substantially changes the above conclusion
 at  $q^2 < 0$.

\vspace{0.4cm}

\subsection{$\Theta(q^2)$ in QCD sum rules}

In ref.\cite{HHMK},
we have evaluated $\lambda$ with the use of the QCD sum rules.
 A main advantage of this method over other estimates
 is that one can relate $\lambda$ to the basic QCD parameters
 such as $m_d/m_u$ and $\la \bar{u}u \ra_0 /\la \bar{d}d \ra_0$.
 For these parameters, we take the results of the chiral perturbation theory
 and QCD sum rules \cite{GLR,HHMK}:
\beq
{{m_d - m_u} \over m_d+m_u}  =  0.28 \pm 0.03 \ , \ \ \ \ \
 {\la \bar{u}u \ra_0 \over \la \bar{d}d \ra_0}  =
 1- (2 \sim 10)\cdot 10^{-3} .
\eeq
After the Borel + finite energy analyses of the QCD sum rules, one obtains
\beq
\lambda = 1.43 \sim  1.85.
\eeq
The corresponding $\Theta(q^2)$ is shown in Fig. 2.

\vspace{6.5cm}

\noindent
Figure 2.  $q^2$ dependence of the non-EM  mixing matrix
 $\Theta$ with $\lambda = [1.43,1.85]$ \cite{HHMK}.

\vspace{0.7cm}

Our  result is  insensitive to the second input
 in eq.(9) which has a large uncertainty.
  Since $\lambda$ is positive and larger than 1,
 $\Theta(q^2)$ in the space like region, which is relevant to N-N force,
 changes sign from its on-shell value.  Thus we establish the
inapplicability of the assumption $\Theta(q^2)=\Theta(m^2)$.

It is in order here to mention other model calculations of $\Theta(q^2)$:
 the constituent-quark loop model ($\rho^0 \rightarrow \bar{q}q \rightarrow
 \omega$) \cite{GHT,KTW} and the nucleon-loop model
 ($\rho^0 \rightarrow \bar{N}N \rightarrow \omega$) \cite{PW}.
 Although they also give a large $q^2$ variation of $\Theta$,
 there is a crucial assumption with no theoretical justification
 in these models.  In fact,  CSB other than
 the EM effect is taken into account
 only as a constituent-quark mass difference or the n-p mass difference,
  while the current-quark mass difference $m_d-m_u$
 is in principle hidden in e.g. the coupling constants of the
 vector mesons with the constituent quarks or the nucleons.
 In the QCD sum rules, one does not suffer from this deficiency, since all
 the CSB effects are automatically taken into account
 in the operator product expansion of the current correlator.
 (For the $\pi^0-\eta$ mixing, the chiral perturbation theory
 is an alternative method to calculate the $q^2$ dependence in a consistent
 way \cite{MAL}.)

To see the physical effect of  $\Theta(q^2)$, let us
 make a Fourier transform of the mixed $\rho^0-\omega$ propagator
 eq.(7)
 and extract a static and central part of the CSB N-N potential, which
 results in
\beq
\bar{V}_{nn} - \bar{V}_{pp} \propto {\Theta(m^2) \over 2m}
 (1- {2\lambda \over mr}) e^{-mr}.
\eeq
$\lambda=0$ corresponds to the potential by Coon and Barrett \cite{CB},
  while $\lambda = 1.78 $ is ours. They are compared  in Fig. 3.
 Although the nn interaction is more attractive than pp when
 $\lambda=0$, the
 use of $\lambda$ larger than 1  completely
 kills the attraction in the relevant range of the CSB  nuclear
 force.  This indicates that the diagram such as Fig. 1(b) alone
 is not effective to  explain the scattering length difference
 $\bar{a}_{nn} - \bar{a}_{pp}$ and other nuclear CSB
 contrary to the previous claims.

\vspace{8cm}

\noindent
Figure 3.  CSB N-N potential $V_{NN}^{\rho \omega} \equiv
 \bar{V}_{nn} - \bar{V}_{pp}$ in an arbitrary unit \cite{HHMK}.
 The solid (dotted) line corresponds to $\lambda=0$ ($\lambda=1.78$).

\vspace{0.7cm}

\section{Outlook --- toward reliable calculations  ---}

We have here analysed a specific CSB effect ``the $\rho^0-\omega$ mixing''
 and found that the effect is substantially changed going from
 the time-like region of $q^2$ to the
 space-like region.  The existence of the $q^2$ dependence
 can be shown in a model independent way using the unsubtracted
 dispersion relation.
 We could also calculate
 the quantitative $q^2$ dependence using the QCD sum rules.
  The resultant CSB potential from the $\rho^0-\omega$ mixing is
  much weaker than the Coon-Barrett's potential and does not seem
 to be a dominant contribution to the various CSB in N-N force
 and in nuclei contrary to the previous claims.

My aim here was not to solve the nuclear CSB puzzles but
  to spotlight a serious problem of the previous approaches
 by taking the $\rho^0-\omega$ mixing as an example.
 One of the major difficulties to study the CSB in many body systems
 lies in the fact that  CSB  is hidden in all over the place
 in hadronic interactions.  As an example, let's take
 a N-N elastic scattering in the one-boson exchange model
 (Fig. 4).

\vspace{7cm}

\noindent
Figure 4.  The N-N scattering with CSB in the one-boson exchange model.
 (a) CSB in the mixing matrix, (b) CSB in external line  and (c)
 CSB in the meson-nucleon vertex.

\vspace{0.7cm}

 The quark mass difference and the EM effect denoted by the ``cross''
  are hidden not only in the  meson mixings (Fig. 4(a)) and
 the external lines (Fig. 4(a)) but also
 in the meson-nucleon vertices (Fig. 4(c)).
 They are all the same order and it is inconsistent to take only a  part
 of them.
   Further difficulties arise from the following facts:
 Firstly, one should consider all possible CSB in scalar (s),
  vector (v),
 pseudoscalar (p) channels in Fig. 2(a,c).
  $q^2$ dependence should be also
 taken into account here.  Secondly,
 it is inconsistent to estimate
 the meson mixings and the CSB meson-nucleon vertices
  by the loop calculations in  hadronic models.
    In fact, there is always CSB effects
 in the tree level of the hadronic effective lagrangians. The
 strength of them cannot be determined in the effective theory
 and must be calculated in QCD or must be fixed by experiments.

These two remarks lead us to the two ways to attach the CSB problem
 in many-body systems: (i) Write down an effective
 lagrangian with all possible CSB terms.
 Calculate the coefficients of these terms
  in QCD (e.g. by the  lattice QCD  or the QCD sum rules).
 Then use the lagrangian to evaluate the relevant hadronic processes.
 (ii) Calculate the physical CSB processes directly in QCD.
  The second approach is  formidable  for many-body systems
 but may be applicable  for N-N force.
 In fact, the studies of the N-N scattering lengths on the lattice
 have been already started \cite{KURA} and
 one may
 generalize the method for the $m_d \neq m_u$ case in the future.

Finally, I should briefly remark on
 the fundamental importance of CSB
  in hadrons and nuclei.  The stability of our world is actually
 intimately related to the $u$-$d$ quark mass difference.
 If it were $m_u > m_d$ just like the quarks in other generations
 (note that $m_c > m_s$ and $ m_t > m_b)$,
 the proton becomes heavier than the neutron
 and  the hydrogens are unstable by the $\beta^+$ decays.
 It will be a fun to see what sort of
 nucleosyntheses takes place in the early universe in such a situation.
 It may be also interesting to note that
 CSB by the EM interaction and that by the quark-mass difference
  are roughly the same order in magnitude
 although they are dictated by independent parameters in the standard model
 ($\alpha=e^2/4\pi$ versus  $(m_d-m_u)/\Lambda_{QCD} =
 (g_d-g_u)\la \phi \ra /\Lambda_{QCD}$  with $g_i$
  being the Higgs-quark coupling and $\la \phi \ra $ being the
 Higgs expectation value). The two effects
 may turn out to have  the same origin in some  unified theory of the
 strong, weak and EM interactions.

\vspace{0.5cm}

\cl{{\bf Acknowledgement}}

I would like to thank the organizing committee for giving me an opportunity
 to talk at this conference. In particular I thank
 Prof. Y. Mizuno and Prof. Toru Suzuki.  This talk is based on a
 work in collaboration with Th. Meissner, E. M. Henley and G. Krein to whom
 I am also grateful.

\end{document}